\documentclass[prb,twocolumn,amsmath,amssymb, letterpaper]{revtex4-1}
\usepackage{amsfonts}
\usepackage{graphicx}
\usepackage{dcolumn}
\usepackage{float}
\usepackage{bm}
\usepackage{mathrsfs}
\usepackage{txfonts}
\usepackage{tabularx}
\usepackage{bigstrut}
\usepackage[usenames,dvipsnames]{color}

\begin{document}

\title{Low-energy theory of a two-dimensional itinerant chiral magnet}

\author{ Armin Rahmani$^1$, Rodrigo A. Muniz$^{1,2,3}$,  and Ivar Martin$^{1,4}$ } 
\affiliation{$^1$Theoretical Division, T4 and CNLS, Los Alamos National Laboratory, Los Alamos, NM 87545, USA \\ 
$^2$International Institute of Physics - UFRN, Natal, RN  59078-400, Brazil \\ 
$^3$Department of Physics and Institute for Optical Sciences, University of Toronto, Toronto, ON, M5S 1A7, Canada \\ 
$^4$Materials Science Division, Argonne National Laboratory, Argonne, IL 60439, USA}

\begin{abstract}
Effective actions, such as nonlinear sigma models, are important tools in studying low-energy fluctuations of magnetically ordered states. Here we derive an effective action for the smooth order-parameter distortions of noncoplanar magnetic textures [with an \textit{SO}(3) order parameter, as opposed to the \textit{SO}$(3)/$\textit{SO}(2)$=S^2$ of collinear magnets] coupled to itinerant electrons. Noncoplanar magnetic textures commonly arise in the presence of electron-mediated long-range spin exchange interactions, giving rise to the emergence of rich electronic phenomena such as anomalous quantum Hall effect. We parameterize the smooth order-parameter distortions of such a magnetic texture by certain non-Abelian fields, and derive the action in terms of these fields by integrating out the itinerant electrons. As a concrete example, we perform explicit calculations for a triangular-lattice model with tetrahedral magnetic ordering. The action encodes the charge and spin quantum-Hall responses as well as the energetics of twisting the magnetic texture.
\end{abstract}

\maketitle

\section{introduction}

Beyond the familiar collinear magnetically ordered states (such as N\'eel antiferromagnets), a myriad of complex magnetic structures can emerge in frustrated magnetic systems even in the classical large-$S$ limit: The next level of complexity arises when the order parameter is coplanar but not collinear (such
as the $120$-degree ordering of a Heisenberg antiferromagent on the triangular lattice). More exotic noncoplanar magnetic textures are in some cases realized by classical magnetic moments with short-range
interactions.~\cite{Momoi1997, Messio2011, Lapa2013} Electron-mediated interactions in itinerant systems (which have a long-range character), on the other hand, commonly give rise to such exotic orders. Indeed, many models of large-$S$ local moments, residing on geometrically frustrated lattices and coupled to itinerant electrons, exhibit energetically stable phases characterized by noncoplanar magnetic textures.~\cite{Martin2008,Akagi2010,Kumar2010,Kato2010,Chern2010,Li2012,Yu2012,Venderbos2012, Chern2011} Due to the presence of a nonvanishing scalar spin chirality, such magnetically ordered itinerant systems may exhibit rich electronic phenomena including spontaneous quantized integer quantum Hall effect~\cite{Ohgushi2000,Shindou2001,Martin2008} and fractionalization on topologically stable defects.~\cite{Rahmani2013}

In addition to specifying the stable magnetic structures, a complete characterization of these exotic phases requires the identification of low-energy magnetic excitations. Such understanding can be obtained from  effective actions for long-wave-length fluctuations around magnetically ordered states [the classic examples being the nonlinear sigma models with (without) topological terms for N\'eel antiferromagnet in one (two) dimensions~\cite{Haldane1983, Affleck1985,Haldane1988b, Ioffe1988,Fradkin1988, Dombre1988, Wen1988}]. Such effective action is not currently available for noncoplanar spin textures stabilized by electron-mediated interactions. In this paper, we take a step toward characterizing the low-energy magnetic excitations, by deriving an effective action for \textit{smooth distortions of the order-parameter}. (In addition to such smooth distortions, noncoplanar magnetic textures may also be distorted by fast fluctuating modes, which are not addressed in the present paper.) The action (i) determines the energetics of smooth distortions, i.e., it allows us to compute the excess energy of any smoothly distorted texture  and hence can be used to evaluate, e.g. the interaction potential between two vortices, and, (ii) encodes the electronic responses to time- and position-dependent perturbations, including the Hall response.

Unlike N\'eel states, where the order parameter can be represented by a unit vector ${\bf m}$, here, the order parameter is specified by a full three-dimensional rotation matrix, which can be parametrized by a unit vector $\bf n$ and a scalar $\phi$ (respectively representing an axis and an angle of rotation of a reference noncoplanar configuration).~\cite{Kawamura1984,Dombre1989} We show in this paper that it is convenient to parameterize the smooth distortions of the order parameter in terms of non-Abelian \textit{SU}$(2)$-gauge-like fields ${\cal A}^\nu_a$ with $\sum_a{\cal A}^\nu_a\sigma_a=-iU^\dagger \partial_\nu U$, $\nu=t,x,y$ (we limit ourselves here to two-dimensional systems) where $\sigma_a$, $a=1\cdots 3$, are the Pauli matrices and $U=\exp\left(-i\phi{\bf n}.{\boldsymbol{\sigma}/2} \right)$. This characterization of spin fluctuations not only encodes the relevant energetics, but also makes the electronic responses of the system transparent. To derive an effective action for the fields above, we need to  integrate out the fermionic degrees of freedom.

To leading order in $\cal A$, the structure of the long-wavelength effective action obtained from the fermionic integration is as follows:
 \begin{equation}\label{eq:struct}
\begin{split}
S_{\rm eff}&=\sum_{{\bf k} }\int_\omega \left[{\cal C}_{ab,\nu \mu}^{0}{\cal A}^\nu_a(k){\cal A}^\mu_b(-k)+i{\cal C}_{ab,\eta\nu \mu}^{1}k_\eta{\cal A}^\nu_a(k){\cal A}^\mu_b(-k)\right],
\end{split}
\end{equation}
where the actual electromagnetic vector potential ${\cal A}^\nu_0$ is treated on the same footing as the the fields ${\cal A}^\nu_a$ parametrizing the distortions for $a>0$. As we show here, the energetics of the distortions are encoded in the mass-term coefficients ${\cal C}_{ab,\nu \mu}^{0}$, while the response-term coefficients ${\cal C}_{ab,\eta\nu \mu}^{1}$ determine the electronic responses of the system.

Focusing on an explicit large-$S$ Kondo-lattice model on the triangular lattice,~\cite{Martin2008} which forms a common nocoplanar texture, known as all-out or tetrahedral,~\cite{Momoi1997},  we explicitly derive the coefficients above, and discuss their implications. Our derivation parallels previous work on the effective action of classical fields coupled to Dirac fermions,~\cite{Volovik1989,Yakovenko1990,Hlousek1990,Abanov2000, Ryu2009} but is done directly on a multisite-unit-cell lattice with nonlinear dispersion. In case of a one-dimensional large-$S$ Kondo lattice model with N\'eel order, a nonlinear sigma model, which also incorporates fast fluctuations, has been derived by integrating out Dirac fermions.~\cite{Tsvelik1994}

The outline of the paper is as follows. In Sec.~\ref{sec:model}, we briefly introduce the model of Ref.~\onlinecite{Martin2008} on the triangular lattice. In Sec.~\ref{sec:action}, we first introduce the non-Abelian fields, which encode the smooth order-parameter distortions of the magnetic medium, and then derive the effective action through explicitly integrating out the fermions. We comment on the physical interpretation of different terms in the action, and, finally, close the paper in Sec.~\ref{sec:conclusion} with a discussion.

\section{model and integer quantum Hall response}
\label{sec:model}

Consider a Kondo-lattice model in the limit of large $S$ ($S\rightarrow \infty$), where the local moments can be treated as classical. Even though there may be a direct interaction between the spins (such as nearest-neighbor Heisenberg), the most novel aspects of the physics of these itinerant systems stem from the electron-mediated interactions, originating from the Kondo coupling of the local moments to itinerant electrons. The Hamiltonian is
\begin{equation}
H=-\sum_{\alpha ij}\left(t_{ij}c_{i\alpha}^{\dagger}c_{j\alpha}+{\rm H.c.}\right)+J\sum_{\alpha\beta i}{\bf S}_{i}\cdot c_{i\alpha}^{\dagger}{\boldsymbol{\sigma}}_{\alpha\beta}c_{i\beta},\label{eq:hamil}
\end{equation}
where $c_{i\alpha}$ is the fermion annihilation operator on site
$i$ with spin $\alpha$, $\boldsymbol{\sigma}$ is a vector of Pauli
matrices, $t_{ij}$ is the electronic hopping between sites $i$ and $j$, and
${\bf S}_{i}$ is a classical magnetic moment on site $i$ (we consider magnetic moments of unit length with their large amplitude $S$ absorbed in the coupling $J$). The magnetic
moments explore different classical configurations with energetics
determined by the quantum fermionic Hamiltonian (the configuration of the magnetic moments can be thought of as external parameters in such Hamiltonians). As a function of the electronic filling and $J/t$, the energetically stable configurations of the local moments determine the magnetic phase diagram of such systems. Although many interesting phases, e.g., stripes, can emerge in such systems,~\cite{Akagi2010} throughout this paper, we focus on noncopanar textures.

The Hamiltonian above may describe a system with two species of itinerant electrons with local spin moments ${\bf S}_i$ as in Kondo lattices, or may alternatively arise in a mean-field decoupling of he Hubbard model, where ${\bf S}_i$ represents the spin-density-wave (SDW) order parameter $\langle c_{i\alpha}^{\dagger}{\boldsymbol{\sigma}}_{\alpha\beta}c_{i\beta}\rangle$. In the latter case as well, the system may exhibit quantum-Hall responses, and similar energetics for smooth distortions of the SDW order parameter.

\begin{figure}[ht]
\includegraphics[width=7cm]{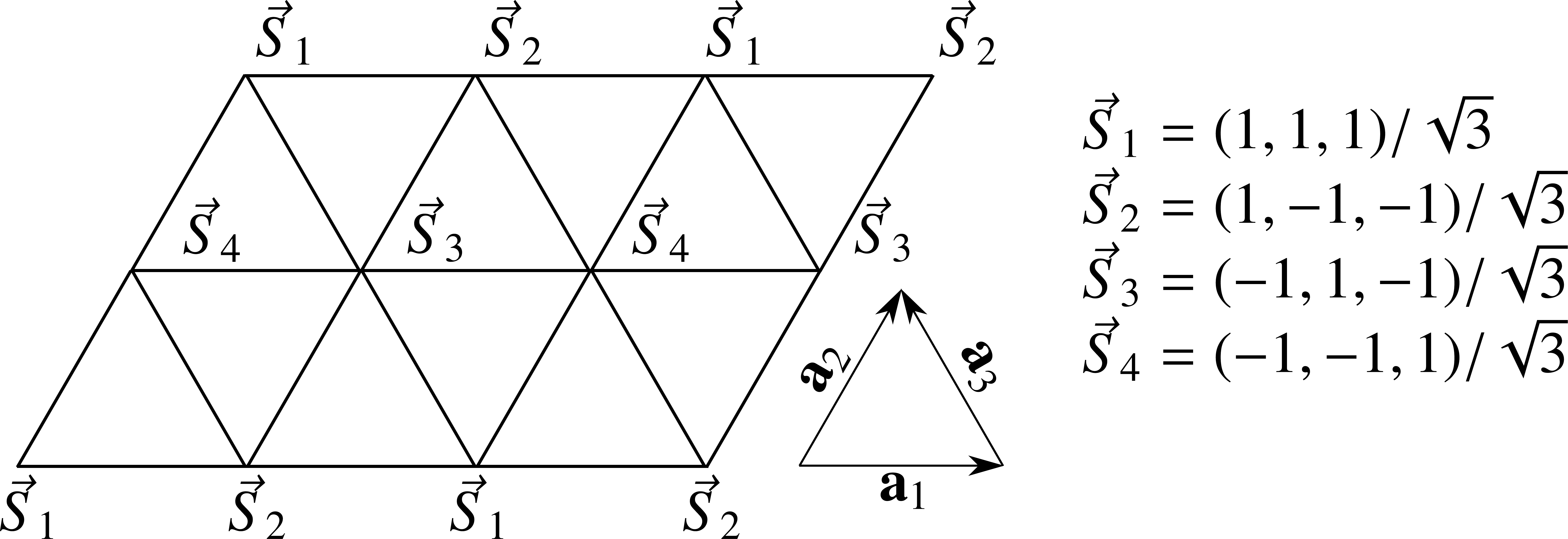} \caption[]{All-out tetrahedral order. For site $i$
in sublattice $a=1\dots4$, ${\mathbf{S}}_{i}=\vec{S}_{a}$. The
four $\vec{S}_{a}$ moments point from the center to the four vertices of a regular tetrahedron. The components are written explicitly in a fixed reference frame. The vectors ${\mathbf{a}}_{i}$ are the lattice vectors.}
\label{fig:lattice} 
\end{figure}

As a concrete example, consider the above Hamiltonian on the triangular lattice with nearest-neighbor electronic hopping. It has been shown that in some regions of the magnetic
phase diagram, the moments ${\bf S}_{i}$ form an all-out tetrahedral noncoplanar
texture, which has a magnetic unit cell consisting of four sites as
shown in Fig.~\ref{fig:lattice}.~\cite{Martin2008,Akagi2010,Kumar2010,Kato2010} We represent the local moments by their components in a fixed Cartesian coordinate system. Note that because there is no spin-orbit coupling in the models we study, this frame is independent of the real-space coordinate system.
Once this texture is stabilized by the electron-mediated interactions (no additional interactions are necessary in this model at $1/4$ and $3/4$ filling fraction), the itinerant electrons experience a nontrivial Berry phase, which results in a gapped integer-quantum-Hall electronic state at precisely the same $1/4$ and $3/4$ filling fractions.

As any noncoplanar structure (in this case the tetrahedral structure formed by four nearby moments on the triangular lattice)  can rotate around any axis by any angle, while preserving the order (and the energy), the order-parameter space is \textit{SO}$(3)$ corresponding to the rotation of a solid object in three dimensions. One can represent the order parameter by a unit vector $\bf n$ (axis of rotation) and a scalar $\phi$ (angle of rotation) of a particular reference state (e.g., the texture shown in Fig.~\ref{fig:lattice} for the triangular lattice model). A global rotation does not change the energy so the effective action should vanish for uniform $\bf n$ and $\phi$. For smooth distortions, a low-energy effective action can be obtained by a gradient expansion in $\partial {\bf n}$ and $\partial \phi$. We emphasize again the distinction of the present problem with the widely studied collinear magnets: in collinear states, the direction of the magnetic moments themselves serves as an order parameter. Since a \textit{SO}$(2)$ rotation around the collinear moment does not change the texture, the order-parameter is the quotient \textit{SO}$(3)/$\textit{SO}(2)$=S^2$ (a two-sphere) instead of full \textit{SO}$(3)$ of our case (see Refs.~\onlinecite{Dombre1989,Kawamura1984})

Before proceeding, let us comment on the limitations of our approach. Our action describes the energetics of smooth distortions of the order parameter, i.e., smooth twisting of the magnetic texture around an slowly changing axis. It describes how such distortions couple to one another as well as to the electromagnetic vector potential in the energy functional of the system, and yields various electronic responses of the system (e.g., charge and spin Hall) to such twists. However, our action does not account for \textit{all} possible fluctuations of the magnetic texture. In case of the triangular-lattice model of Fig.~\ref{fig:lattice}, the unit cell has four sublattices, and each local moment can point in any direction (characterized by two angles). So the total number of modes for a unit cell is equal to eight. The \textit{SO}$(3)$ order parameter is only characterized by three numbers so in addition to the space- and time-dependent orientation of the tetrahedral order, which changes smoothly to remain in the low-energy sector, there are five gapped modes that derive from the intra-unit cell magnetic distortions. Inclusion of these gapped modes can be done in analogy to the inclusion of the ferromagnetic fluctuations in the N\'eel order case~\cite{Fradkin1991,Wen2004}. A similar issue emerges also in the case of 120-degree order on the triangular lattice,~\cite{Dombre1989} where overall magnetization fluctuations are included in addition to the \textit{SO}(3) order parameter. A full theory of low-energy magnetic fluctuations in our case(tetrahedral magnetic structure in the triangular lattice) must also include the five gapped modes in addition to the order-parameter distortions. In analogy with Ref.~\onlinecite{Dombre1989}, we expect that such fast modes only couple to the smooth order-parameter distortions in the Berry-phase terms and not in the energy functional. As such, they can affect the dynamics but not the energetics of smoothly twisting the magnetic texture or the electronic responses. Similarly, deriving a full nonlinear sigma model for noncoplanar textures remains and open problem.

\section{effective action for distortions}
\label{sec:action} 

In this section, we derive an effective action for smooth distortions of the magnetic medium around the all-out state of Fig.~\ref{fig:lattice}. First, we argue that certain non-Abelian gauge-like fields provide a convenient set of parameters for writing out the effective action. We then derive the action in terms of these fields by explicitly integrating out the fermions as discussed in the proceeding subsections.

\subsection{Parameterizing the distortions}
In the absence of an electromagnetic external gauge field, the Hamiltonian~\eqref{eq:hamil} leads to the following action:
\begin {equation}\label{eq:S_0}
\begin{split}
S=\int d\tau\sum_{\mathbf r}\Bigg\lbrace&\overline{\psi}({\mathbf r},\tau) \partial_\tau \psi ({\mathbf r},\tau) +\sum_{i} \Big[-t \overline{\psi}({\mathbf r},\tau)  \psi ({\mathbf r}+{\mathbf a}_i,\tau) \\
&+{\rm H.c.}\Big]
+J\: \overline{\psi}({\mathbf r},\tau) \left[R({\mathbf r},\tau ){\mathbf S}_0({\mathbf r} )\cdot {\boldsymbol\sigma}\right]\psi ({\mathbf r},\tau) \Bigg\rbrace,
\end{split}
\end{equation}
where $\psi$ is a two-component (for spin up and down) Grassman variable and $R$ is a smooth \textit{SO}$(3)$ rotation matrix related to $\bf n$ and $\phi$ through
\begin {equation}\label{eq:R}
R=\cos \phi \openone+\sin \phi [{\bf n}]_\times+(1-\cos \phi) {\bf n}\otimes{\bf n}^T,
\end{equation}
where $[{\bf n}]_\times$ is the cross-product matrix of the unit vector $\bf n$, which yields the cross product ${\bf n}\times {\bf x}=[{\bf n}]_\times{\bf x}$ when acting on any vector $\bf x$, and ${\bf n}\otimes{\bf n}^T$ is the projection matrix of the $\bf n$ direction. Both $\bf n$ and $\phi$ are smooth functions of $\bf r$ and $\tau$. The lattice vectors ${\bf a}_i$ represent the the bonds connecting the nearest-neighbor sites (without loss of generality, let us consider the triangular lattice model where $i=1\dots 3$) and the moments ${\bf S}_0({\bf r})$ correspond to the noncopanar texture stabilized in the magnetic phase under study. Note that although the rotation matrix is defined on each site (as opposed to unit cell), the constraint of smoothness of the rotation matrix restricts the possible spin orientations to smooth variations of the order parameter. In case of the triangular-lattice example, the magnetic moments ${\mathbf S}_0({\mathbf r})$ take on the values shown in Fig.~\ref{fig:lattice} on the four sublattices $a=1\cdots 4$  in the tetrahedral phase. The three lattice vectors ${\mathbf a}_i$ are also shown in Fig.~\ref{fig:lattice} for the triangular-lattice model. On each lattice site, the rotation matrix $R$, which characterizes the distortions of a reference ordered state, is described by three independent real parameters (two for $\bf n$ and one for $\phi$).

We now make a change of Grassman variable using an \textit{SU}$(2)$ transformation $U=\exp\left(-i\phi{\bf n}.{\boldsymbol{\sigma}/2} \right)$: $\psi=U \chi$.  The change of variable is chosen so as to absorb the rotation of the magnetic moments in the definition of new fermions $\chi$.~\cite{Rahmani2013} In other words, we have 
\begin {equation}
U({\mathbf r},\tau){\mathbf S}_0({\mathbf r} )\cdot {\boldsymbol\sigma} U^\dagger({\mathbf r},\tau)=\left[ R({\mathbf r},\tau){\mathbf S}_0({\mathbf r} )\right]\cdot {\boldsymbol\sigma},
\end{equation}
which leads to a term in the Lagrangian of the form $J\: \overline{\chi} \left[{\mathbf S}_0\cdot {\boldsymbol\sigma}\right]\chi$. The effect of the distortions now appears in the hopping terms in the action. We can then write the action as 
\begin{equation}\label{eq:split}
S=S_0+S_{{\mathbf A}},
\end{equation}
where $S_0$ is the action of a uniform noncoplanar texture in the absence of the distortions characterized by $R$ ($S_0$ has the same form as Eq.~\eqref{eq:S_0} with the substitution $R\rightarrow1$ and $\psi\rightarrow\chi$),
and $S_{\mathbf A}$ comes from the inserting $\psi=U \chi$ into the hopping terms. Notice that just like the $3\times3$ rotation matrix $R$, the $2\times2$ matrix $U$ is also characterized by three real parameters.

Let us now write out $S_{{\mathbf A}}$ corresponding to nonuniform distortions. From the first term in Eq.~\eqref{eq:S_0}, we get the contribution $\overline{\chi}({\mathbf r},\tau)\left({U}^\dagger \partial_\tau{U} \right) \chi ({\mathbf r},\tau)$ to the Lagrangian ${\cal L}_A$. 
%Note that the imaginary time $\tau$ is purely imaginary and thus all ${\cal A}^\tau_\mu$ are real due to the unitarity of $U$.
Similarly, the second term in Eq.~\eqref{eq:S_0} gives contributions of the form $-t\overline{\chi}({\mathbf r},\tau)\left[{U}^\dagger({\mathbf r},\tau) {U}({\mathbf r}+{\mathbf a}_i,\tau)  \right]\chi ({\mathbf r}+{\mathbf a}_i,\tau)+{\rm H.c.}$. For smooth $U$ (over length scales of the order of the lattice spacing), we can expand the above term in gradients. Up to second order, we have
\begin{equation}\label{eq:expand}
\begin{split}
{U}^\dagger &({\mathbf r}) {U}({\mathbf r}+{\mathbf d})\approx \left[{U}^\dagger-{d_i\over 2}\:\partial_i{U}^\dagger+{d_i d_j\over 8}\:\partial_i\partial_j U ^\dagger\right]\\
&\times\left[ U +{d_{i'}\over 2}\:\partial_{i'} U +{d_{i'} d_{j'}\over 8}\:\partial_{i'}\partial_{j'} U\right]\bigg{|}_{{\mathbf r}+{{\mathbf d}/2}},
\\
%&\approx \bigg {\{}1+{d_i\over 2}\:\left[U^\dagger\partial_i U-(\partial_i U^\dagger)U\right]+{d_i d_j\over 8}\times\\
%&\:\left[-2\left(\partial_i U^\dagger\right)\partial_j U+U^\dagger\partial_i\partial_j U+\left(\partial_i\partial_j U^\dagger\right)U \right]\bigg {\}}\bigg{|}_{{\mathbf r}+{{\mathbf d}\over 2}},
\end {split}
\end{equation}
where summation over repeated indices is implied.

Since $U^\dagger U=1$, we can then write upon differentiation:
\begin{eqnarray}
0 &=& U^\dagger\partial_i U+(\partial_i U^\dagger)U,\\
0 &=&\left(\partial_i U^\dagger\right)\partial_j U+\left(\partial_j U^\dagger\right)\partial_i U+U^\dagger\partial_i\partial_j U+\left(\partial_i\partial_j U^\dagger\right)U,\nonumber
\end{eqnarray}
which, after substituting in Eq.\eqref{eq:expand}, leads to
\begin{equation}\label{eq:expand2}
{U}^\dagger({\mathbf r}) {U}({\mathbf r}+{\mathbf d})\approx
1+d_i\:U^\dagger\partial_i U+
{d_i d_j\over 2}\left(U^\dagger \partial_i U \right)\left(U^\dagger \partial_j U \right),
\end{equation}
with the right-hand side computed at ${\bf r}+{\bf d}/2$. Notice that due to the unitarity of $U$, the second derivatives are related to the first ones and the second-order expansion~\eqref{eq:expand2} can be written in terms of quantities $U^\dagger\partial_\mu U$ ($\mu=\tau,x,y$). We thus find that the matrices $U^\dagger\partial_\mu U$ provide a convenient parameterization of the smooth distortions of a uniform noncoplanar magnetic texture (such as the tetrahedral texture of Fig.~\ref{fig:lattice} on the triangular lattice). With a little algebra, we can show that $iU^\dagger\partial_\mu U$ is related to $\bf n$ and $\phi$ and their derivatives through
\begin{equation}\label{eq:A_n_phi}
\begin{split}
i U^\dagger \partial_\mu U=&({\bf n}\cdot {\boldsymbol\sigma})\partial_\mu \phi+{\sin \phi \over 2}\left(\partial_\mu {\bf n}\cdot {\boldsymbol \sigma}\right)\\
&+{\cos \phi-1\over 2}\left[\left({\bf n}\times{\partial_\mu{\bf n}}\right)\cdot{\boldsymbol \sigma}\right].
\end{split}
\end{equation}
Therefore deriving an effective action for $U^\dagger \partial_\mu U$ immediately yields an action for the original geometric fields ${\bf n}$ and $\phi$ through substituting Eq.~\eqref{eq:A_n_phi}.

Let us define the coefficients of the above expansion in Pauli matrices as
\begin{equation}\label{eq:A_def}
-U^\dagger \partial _\tau U={\cal A}^\tau_a \sigma^a, \quad -i U^\dagger \nabla U={\cal A}^{\mathbf r}_a \sigma^a, \quad a=1\dots 3,
\end{equation}
where summation over $a$ is implicit. In this notation, the electromagnetic $U(1)$ gauge fields ${\cal A}^\mu_0$ can be added on the same footing (even though we do not have full \textit{SU}$(2)$ symmetry) as the gauge-like fields ${\cal A}^\mu_a$ for $a>0$, which were constructed in order to characterize the distortions of the noncoplanar magnetic medium. It is important to distinguish the origin of the fields with $a=0$ and $a>0$: while the former is a real gauge field with a well-defined $U(1)$ gauge transformation, the latter merely resembles a gauge field. The fields ${\cal A}^\mu_a$ for $a>0$ do not have their own dynamics (Maxwell term) and a gauge structure (in fact if we did have a gauge structure, these fields would be pure gauge and could be gauged away); they are just some parameters characterizing the smooth distortion of the magnetic medium around a reference noncoplanar texture. For ${\cal A}^\mu_0$, on the other hand, we have a Maxwell term determining the dynamics, and the standard $U(1)$ gauge symmetry.

 Putting Eqs.~\eqref{eq:A_def} and ~\eqref{eq:expand2} together, we can finally write $S_{\mathbf A}$ [see Eq.~\eqref{eq:split}] as
\begin {equation}\label{eq:s_a0}
\begin{split}
S_{\mathbf A}=&-\int d \tau \sum_{\mathbf r}\:\Bigg\lbrace\overline{\chi}({\mathbf r}){\mathbf A}^\tau({\mathbf r}) \chi ({\mathbf r})\\
& +t \sum_{j=1}^3\:\left[\overline{\chi}({\mathbf r}) {\cal K}({\bf r},{\bf a}_j)\chi ({\mathbf r}+{\mathbf a}_j) +{\rm H.c.}\right]
 \Bigg\rbrace,
 \end{split}
\end{equation}
where 
\begin {equation}\label{eq:K}
\begin{split}
&{\mathbf A}^\mu\equiv {\cal A}^\mu_\nu \sigma^\nu, \quad \nu=0\dots 3,\\
&{\cal K}({\bf r},{\bf a}_j)\equiv i\:{\mathbf a}_j\cdot {\mathbf A}^ {\mathbf r} ({\mathbf r} +{{\mathbf a}_j / 2})-{1 \over 2} \left[{\mathbf a}_j\cdot {\mathbf A}^ {\mathbf r} ({\mathbf r} +{{\mathbf a}_j / 2})\right]^2.
 \end{split}
\end{equation}
Note that $\sigma^0$ represents the $2\times 2$ identity matrix. Also notice that the explicit dependence of ${\mathbf A}$ and $\chi$ in the above expressions all on $\tau$ is suppressed for brevity. 

In this paper, we discuss how an effective action for ${\cal A}_\nu^\mu$ can be derived by
\begin{enumerate}
\item Fourier transforming the action.
\item Integrating out the fermionic degrees of freedom.
\item Expanding the resulting action in ${\cal A}_\nu^\mu$ (justified for smooth distortions due to the presence of derivatives in the definition of the ${\cal A}_\nu^\mu$ fields).
\item Expanding in momenta as we are interested in the long-distance low-energy behavior.
\end{enumerate}
We will carry out the above procedure explicitly for the triangular-lattice model. However, as can be seen from the generic structure, the form~\eqref{eq:struct} and the physical implication of the effective action are not specific to this model. The value of certain nonuniversal coefficients, which we compute explicitly, however, are special properties of the triangular-lattice model.

\subsection {Full action in momentum space}

In order to integrate out the fermions, it is helpful to write the action in momentum space. If the unit cell of the noncoplanar structure of interest has $M$ sites, the Hamiltonian in momentum space can be simply written as a $2M\times 2M$ matrix, where the factor of $2$ accounts for the two spin species. The diagonal elements of the Hamiltonian are given by $\vec{S}_i\cdot{\boldsymbol\sigma} $, where $\vec{S}_i$ is the local moment on sublattice $i$ in a reference texture. The off-diagonal hopping terms are given by $-2 t \cos ( {\mathbf k}\cdot {\mathbf a}_{ij})\sigma^0$, where ${\mathbf a}_{ij}$ is the lattice vector connecting sublattice $i$ to sublattice $j$. In case of the triangular lattice model of Fig.~\ref{fig:lattice}, we have the following Hamiltonian
\begin{equation}\label{eq:H+momentum}
{\cal H}({\mathbf k})=J
\left(\begin{array}{cccc}
\vec{S}_1\cdot{\boldsymbol\sigma} &
0 &
 0 &
  0
   \\
0&
\vec{S}_2\cdot{\boldsymbol\sigma} &
0&
0
\\
0&
0 &
\vec{S}_3\cdot{\boldsymbol\sigma} &
0
\\
0&
0 &
0&
\vec{S}_4\cdot{\boldsymbol\sigma} 
\end{array} 
\right)+
{\cal E}({\mathbf k})\otimes \sigma^0,
\end{equation}
where the $4\times 4$ matrix ${\cal E}({\mathbf k})$ is given by 
\begin{equation}\label{eq:E_matrix}
{\cal E}({\mathbf k})=
\left(\begin{array}{cccc}
0 &
 \epsilon_1({\mathbf k})&
 \epsilon_3({\mathbf k})  &
 \epsilon_2({\mathbf k})
   \\
 \epsilon_1({\mathbf k})&
0 &
 \epsilon_2({\mathbf k})&
 \epsilon_3({\mathbf k})
\\
  \epsilon_3({\mathbf k})&
 \epsilon_2({\mathbf k}) &
0 &
 \epsilon_1({\mathbf k})
\\
 \epsilon_2({\mathbf k})&
 \epsilon_3({\mathbf k} )&
 \epsilon_1({\mathbf k}) &
0
\end{array} 
\right),\quad \epsilon_i({\mathbf k})\equiv -2 t \cos ( {\mathbf k}\cdot {\mathbf a}_i),
\end{equation}
in the basis given by the eight-component vector
\begin{equation}
{\Psi}^\dagger({\mathbf r})=\Big(\chi_1^\dagger({\mathbf r}),\chi_2^\dagger({\mathbf r}+{\mathbf a}_1),\chi_3^\dagger({\mathbf r}+{\mathbf a}_1+{\mathbf a}_2),\chi^\dagger_4({\mathbf r}+{\mathbf a}_2)\Big).
\end{equation}
A similar structure arises in the general case for a $2M$-component ${\Psi}^\dagger({\mathbf r})$ and an $M\times M$ matrix ${\cal E}({\mathbf k})$.

To write the action, we replace the above creation operators (and similarly for the annihilation operators) by Grassman variables and endow each Grassman variable with time-dependence, i.e., ${\Psi}^\dagger({\mathbf r})\rightarrow \overline{\Psi}({\mathbf r}, \tau)$.
Upon Fourier transformation, we can now write $S_0$ as
%\begin{equation}
%S_0= \sum_{{\mathbf k}}\int_\omega\:\left[-i\omega\overline{\Psi}({\mathbf k},\omega)\Psi ({\mathbf k},\omega)+ \overline{\Psi}({\mathbf k},\omega) {\cal H}({\mathbf k})\Psi ({\mathbf k},\omega) \right],
%\end{equation}
\begin{equation}
S_0= -\sum_{{\mathbf k}}\int_\omega \overline{\Psi}({\mathbf k},\omega)G_0^{-1}({\mathbf k},\omega)\Psi ({\mathbf k},\omega),
\end{equation}
where the inverse Green's function $G_0^{-1}({\mathbf k},\omega)=i\omega-{\cal H}({\mathbf k})$ and the $2M\times 2M$ ($8\times 8$ in case of the triangular lattice model of  Fig.~\ref{fig:lattice}) matrix ${\cal H}({\mathbf k})$ represents the momentum-space Hamiltonian matrix [see Eq.~\eqref{eq:H+momentum}].
 We can now similarly write $S_{\mathbf A}$ [see Eq.~\eqref{eq:s_a0}] in momentum space. Setting factors of volume to unity, we have
$
S_{\mathbf A}=-\sum_{{\mathbf k}_1,{\mathbf k}_2}\int_{\omega_1,\omega_2}
\overline{\Psi} ({\mathbf k}_1,\omega_1)B({\bf k}_1, {\bf k}_2, \omega_1-\omega_2)\Psi ({\mathbf k}_2,\omega_2)$, with
 \begin{widetext}
 \begin {equation}\label{eq:S_A1}\begin{split}
B({\bf k}_1, {\bf k}_2, \omega)&\equiv
{\cal A}^\tau_a({\mathbf k}_1-{\mathbf k}_2,\omega) \openone_M \otimes\sigma ^a 
-{\cal A}^{ i}_a({\mathbf k}_1-{\mathbf k}_2,\omega)
\left( \partial_{k_i} {\cal E}({\mathbf k}) \otimes \sigma^a\right)\Big |_{\left({{\mathbf k}_1+{\mathbf k}_2\over 2}\right)}\\
&+{1\over 2}\sum_{{\bf k}_3, \omega_3}{\cal A}^{ i}_a({\mathbf k}_3,\omega_3){\cal A}^{ j}_b({\mathbf k}_1-{\mathbf k}_2-{\mathbf k}_3,\omega-\omega_3)\left[ \partial_{k_i}  \partial_{k_j}{\cal E}({\mathbf k}) \otimes \left(\sigma^a\sigma^b\right)\right]\Big |_{\left({{\mathbf k}_1+{\mathbf k}_2\over 2}\right)},
\end{split}
\end{equation}
\end{widetext}
where $\openone_M$ is the $M\times M$ identity matrix.
In the above expression, summation over repeated indices is implied ($i$ and $j$ are summed over $x, y$ and $a$ and $b$ over $0\dots 3$). Notice that unless one of the $a$ or $b$ indices is equal to zero, the last term vanishes for $a\neq b$ as the two different Pauli matrices anticommute.

\subsection{Integrating out fermions}\label{sec:int}
It is well-known that for Grassman variables $\varsigma_i$ and a matrix $\L$, we have $\int \prod_i d \overline{\varsigma}_id\varsigma_i e^{-\sum_{ij}\overline{\varsigma}_i L_{ij}\varsigma_j}=\det L$. Treating the momentum and frequency dependence of the fields $\Psi$ as matrix indices, performing the fermionic path integral over $\Psi$, and using the matrix identity $\ln\left( \det L\right)={\rm tr}\left( \ln L\right)$ leads to the following exact effective action
\begin{equation}
S_{\rm eff}=-{\rm Tr} \left[\ln\left({G}_0^{-1}+{ B}\right)\right]=S_0-{\rm Tr} \left[\ln\left(\openone+{ G}_0{ B}\right)\right],
\end{equation}
%
%We now perform the fermionic path integral over $\Theta$ and write the effective action as
%\begin{equation}
%S_{\rm eff}=-{\rm Tr} \left[\ln\left({\cal G}_0^{-1}+{\cal B}\right)\right]=S_0-{\rm Tr} \left[\ln\left(\openone+{\cal G}_0{\cal B}\right)\right],
%\end{equation}
where ${\rm Tr}$ indicates a trace over the $2M$ matrix indices as well as $\mathbf k$ and $\omega$ (we shall represent a trace over only the $2M$ matrix indices by ``tr''). Notice that the Green's function $G_0$ is diagonal in the $\bf k$ and $\omega$ indices. As mentioned before, in the limit of smooth distortions, the fields ${\cal A}_\nu^\mu$ are small due to the derivatives in their definitions. Therefore, we can compute the effective action through an expansion in powers of ${\cal A}_\nu^\mu$, and truncating the expansion at a given order. Here, we go up to second order.

To obtain such expansion in ${\cal A}_\nu^\mu$, we observe that the matrix $B$ [see Eq.~\eqref{eq:S_A1}] is comprised of first and second order terms in ${\cal A}_\nu^\mu$:
\begin{equation}
B=B^{(1)}+B^{(2)}.
\end{equation}
Using the above and the following expansion of the logarithm $\ln\left(\openone+{G}_0{ B}\right)={ G}_0{B}-\left({\cal G}_0{ B}\right)^2/2+\left({G}_0{ B}\right)^3/3+\dots$, we obtain
\begin{eqnarray}
S_{\rm eff}^{(1)}&=&-{\rm Tr}\left(G_0B^{(1)}\right),\label{eq:s1}\\
S_{\rm eff}^{(2)}&=&-{\rm Tr}\left(G_0B^{(2)}\right)+{1\over 2}{\rm Tr}\left(G_0B^{(1)}G_0B^{(1)}\right),\label{eq:s2}
%S_{\rm eff}^{(3)}&=&{\rm Tr}\left(G_0B^{(1)}G_0B^{(2)}\right)-{1\over 3}\left(G_0B^{(1)}G_0B^{(1)}G_0B^{(1)}\right),\label{eq:s3}
\end{eqnarray}
where the first (second) term on the right-hand side of Eq.~\eqref{eq:s2} comes from the first (second) order term in the expansion of the logarithm. 
%We have made use of the cyclic property of the trace to write ${1\over 2}{\rm Tr}\left(G_0B^{(1)}G_0B^{(2)}\right)+{1\over 2}{\rm Tr}\left(G_0B^{(2)}G_0B^{(1)}\right)={\rm Tr}\left(G_0B^{(1)}G_0B^{(2)}\right)$ in Eq.~\eqref{eq:s3}.
The equations above provide the starting point of the calculation of the effective action.

 Before focusing on the triangular lattice case, where we can disentangle the spin and subsystem indices due to a symmetry, we comment on the general structure of the calculation through an example. Let us first simplify the notation by introducing $k=(k_0,k_1,k_2)$, where $k_0\equiv i \omega$ and $k_{1,2}\equiv k_{x,y}$, and the following shorthand notation:
\begin{equation}
\partial_0=-i\partial_\omega,\qquad \partial_1=\partial_{k_x},\qquad \partial_2=\partial_{k_y}.
\end{equation}
We now define operators $J_\mu^a({\bf k})\equiv \partial_\mu \left[ i\omega-{\cal E}({\bf k}) \right]\otimes\sigma^a$, such that from Eq.~\eqref{eq:S_A1}, we have 
\begin{equation}\label{eq:J}
B^{(1)}({\bf k}_1,{\bf k}_2, \omega_1-\omega_2)={\cal A}_a^\mu({\bf k}_1-{\bf k}_2, \omega_1-\omega_2)J_\mu^a\left({{\bf k}_1+{\bf k}_2\over 2}\right).
\end{equation}
Notice that $J_0^a({\bf k})=\openone_M\otimes \sigma^a$ is independent of $k$.

As a concrete example of the computations involved, let us consider the term ${\rm Tr}\left(G_0B^{(1)}G_0B^{(1)}\right)$, which appears in $S_{\rm eff}^{(2)}$ [see Eq.~\eqref{eq:s2}]. Each $G_0$ or $B^{(1)}$ is, in the most general case, labeled by two indices $k_i, k_j$. As $G_0$ is diagonal in these indices, we need to compute an integral over $k_1$ and $k_2$ of the following trace of a $2M\times 2M$ matrix: ${\rm tr}\left[G_0(k_1)B^{(1)}(k_1,k_2)G_0(k_2)B^{(1)}(k_2,k_1)\right]$. The frequency $\omega$ in such integration runs from $-\infty$ to $+\infty$ and the sum over discrete momenta $\bf k$ reduces to an integral over the Brillouin zone in the thermodynamic limit.
Inserting Eq.~\eqref{eq:J} into the above expression gives ${\cal A}^\mu_a (k_1-k_2){\cal A}^\nu_b (k_2-k_1){\rm tr}\left[G_0(k_1)J^a_\mu(k_1,k_2)G_0(k_2)J^b_\nu(k_2,k_1)\right]$.

To write an action for $\cal A$, it is natural to make a change of variables $k_1-k_2=k$ and $k_2=k'$. The trace multiplying ${\cal A}^\mu_a (k){\cal A}^\nu_b (-k)$ can then be expanded in $k$ (as we are interested in a long-wavelength action). The main calculation then involves an integration over $k'$ to find the coefficients. Such calculation can be performed by diagonalizing the Hamiltonian ${\cal H}({\bf k})$ so that the Green's function is written as $G_0({\bf k},\omega)=\sum_m {1\over i\omega-E_m({\bf k})}|m({\bf k})\rangle \langle m({\bf k})|$, where $E_m({\bf k})$ and $|m({\bf k})\rangle$ are the eigenvalues and eigenvectors of the Hamiltonian. The integral over $\omega$ can then be exactly performed by contour integration, leaving us with just a finite momentum integral over the Brillouin zone. It then turns out that the general structure of the effective action, to the leading order we study, is given by Eq.~\eqref{eq:struct}. In the following section, we explicitly compute the coefficients in Eq.~\eqref{eq:struct} for the triangular lattice model of Fig.~\ref{fig:lattice} with tetrahedral ordering, and discuss their implications.

%
% We are interested in an expansion in powers of $\cal A$ of $S_{\rm eff}$, in the long-wavelength low-frequency limit. The logarithm can be expanded as $\ln\left(\openone+{\cal G}_0{\cal B}\right)=+{\cal G}_0{\cal B}-\left({\cal G}_0{\cal B}\right)^2/2+\left({\cal G}_0{\cal B}\right)^3/3+\dots$, while $\cal B$ itself has terms of first and second order in $\cal A$: ${\cal B}={\cal B}^{(1)}+{\cal B}^{(2)}$.

\subsection{Explicit calculation of the coefficients on the triangular lattice}

The noncoplanar texture of Fig.~\ref{fig:lattice} has a symmetry (a combination of lattice translation and spin rotation),~\cite{Martin2008} which allows us to write the $8\times 8$ Hamiltonian in a block-diagonal form (with two identical locks) using a unitary transformation $\cal U$ [see Eq.~\eqref{eq:H+momentum} and Appendix.~\ref{app1}]:
\begin{equation}\label{eq:unit}
\Theta={\cal U}\Psi,\qquad \Psi^\dagger({\bf k}) {\cal H}({\bf k}) \Psi({\bf k}) =\Theta^\dagger({\bf k})\left[ \sigma^0\otimes{\mathscr H}({\bf k}) \right]\Theta({\bf k}).
\end{equation}
This transformation allows us to write the bare action $S_0$ as 
\begin{equation}
S_0=- \sum_{{\mathbf k}}\int_\omega\overline{\Theta}({\mathbf k},\omega)\left[\sigma^0\otimes{\mathscr G}^{-1}({\mathbf k}, \omega)\right]\Theta({\mathbf k},\omega),
\end{equation}
where 
\begin{equation}\label{eq:greens}
{\mathscr G}^{-1}({\mathbf k}, \omega)\equiv i \omega -{\mathscr H}({\mathbf k}).
\end{equation}
Interestingly, one can check that, for all Pauli matrices $\sigma^a$, the same transformation acting on ${\cal E}({\bf k})\otimes\sigma^a$ and $\openone \otimes \sigma^a$ [see Eq.~\eqref{eq:S_A1}] gives
\begin{eqnarray}
{\cal U}\left[{\cal E}({\bf k})\otimes\sigma^a\right]{\cal U}^\dagger&=&\sigma^a\otimes\left[{\mathscr H}({\bf k}) {\cal D}_a\right]\big{|}_{J=0},\label{eq:prop1}\\
{\cal U}\left[\openone\otimes\sigma^a\right]{\cal U}^\dagger&=&\sigma^a\otimes {\cal D}_a,\label{eq:prop2}
\end{eqnarray}
where the diagonal matrices ${\cal D}_a$ are defined as
\begin{eqnarray*}
{\cal D}_0 & \equiv & \sigma^0 \otimes \sigma^0,\qquad
{\cal D}_1  \equiv  \sigma^0 \otimes \sigma^3, \\
{\cal D}_2& \equiv & \sigma^3\otimes \sigma^0, \qquad
{\cal D}_3\equiv  \sigma^3\otimes \sigma^3.
\end{eqnarray*}
We can then cast Eq.~\eqref{eq:S_A1} to the following form:
\begin{equation}
S_{\mathbf A}=-\sum_{{\mathbf k}_1,{\mathbf k}_2}\int_{\omega_1,\omega_2}\overline{\Theta} ({\mathbf k}_1,\omega_1){\cal B}({\mathbf k}_1,{\mathbf k}_2,\omega_1-\omega_2)\Theta ({\mathbf k}_2,\omega_2),
\end{equation}
with ${\cal B}({\mathbf k}_1,{\mathbf k}_2,\omega)={\cal B}^{(1)}({\mathbf k}_1,{\mathbf k}_2,\omega)+{\cal B}^{(2)}({\mathbf k}_1,{\mathbf k}_2,\omega)$:
\begin{equation}\label{eq:Bs}
\begin{split}
&{\cal B}^{(1)}({\mathbf k}_1,{\mathbf k}_2,\omega)=
{\cal A}^{ \mu}_a({\mathbf k}_1-{\mathbf k}_2,\omega) \:\sigma^a\otimes
\left[\partial_{\mu} {\mathscr G}^{-1}\left({{\mathbf k}_1+{\mathbf k}_2\over 2}\right)  {\cal D}_a\right],\\
&{\cal B}^{(2)}({\mathbf k}_1,{\mathbf k}_2,\omega)=-{1 \over 2} \sum_{{\bf k}'}\int_{\omega'}{\cal A}^{\mu}_a({\mathbf k}',\omega'){\cal A}^{ \nu}_b({\mathbf k}_1-{\mathbf k}_2-{\mathbf k}',\omega-\omega')\\
&\times  \left(\sigma^a\sigma^b\right)\otimes\left[\partial_{\mu}  \partial_{\nu} {\mathscr G}^{-1}\left({{\mathbf k}_1+{\mathbf k}_2\over 2}\right)  {\cal D}_a{\cal D}_b\right].
\end{split}
\end{equation}
Notice that $\partial_\nu {\mathscr G}^{-1}$ for $\nu=0,1,2$ is independent of $\omega$.

\subsubsection{First order in ${\cal A}$}
As a warmup, let us start by the first-order term, which comes from Eq.~\eqref{eq:s1} with the substitution $B^{(1)}\rightarrow {\cal B}^{(1)}$ [see Eq.~\eqref{eq:Bs}] and $G_0\rightarrow \sigma^0\otimes {\mathscr G}$ [see Eq.~\eqref{eq:greens}] due to the transformation~\eqref{eq:unit}:
\begin{equation}\label{eq:s_eff1}
S^{(1)}_{\rm eff}=-{\rm tr}\left(\sigma^a\right) {\cal A}^\nu_a(0,0)
\sum_{\bf k} \int_\omega {\rm tr}\left[{\mathscr G}({\bf k},\omega)\partial_\nu {\mathscr G}^{-1}({\bf k}){\cal D}_a \right].
\end{equation}
 In deriving the above expression we have made use of the following trace identity:
\begin{equation}\label{eq:trace}
{\rm tr}\left[\left(A_1\otimes B_1\right)\dots \left(A_n\otimes B_n\right)\right]={\rm tr}\left(A_1\dots A_n\right){\rm tr}\left(B_1\dots B_n\right).
\end{equation}
The above expression for for $a>0$ obviously vanishes due to the zero traces of the Pauli matrices. However, for $a=0$, as discussed in Appendix.~\ref{app_b1}, we get 
\begin{equation}
S^{(1)}_{\rm eff}= \pi {\cal A}^0_0(0,0)\sum_{n,{\bf k}}\left(2n_0(\varepsilon^n_{\bf k})-1\right).
\end{equation}
This expression simply corresponds to a chemical potential proportional to the scalar electromagnetic potential ${\cal A}^0_0(0,0)$, which is not relevant to the distortion of the magnetic structure and therefore not included in Eq.~\eqref{eq:struct}.

\subsubsection{Second order in ${\cal A}$}
As for the second-order term $S^{(2)}_{\rm eff}$, we have two contributions according to Eq.~\eqref{eq:s2}: one from ${\cal B}^{(1)}$ at second order in the expansion of the logarithm, $S^{(2),(1)}_{\rm eff}$, and one from ${\cal B}^{(2)}$ at first order, $S^{(2),(2)}_{\rm eff}$. Using our general method, we can write these terms as follows: 
\begin{eqnarray}
& &\begin{split}
S^{(2),(1)}_{\rm eff}=&{1 \over 2}{\rm tr}\left(\sigma^a \sigma^b\right)\sum_{{\bf k}_1 {\bf k}_2}\int _{\omega_1,\omega_2}
{\cal A}^\nu_a(k_1-k_2)
{\cal A}^\mu_b(k_2-k_1)\\
& \times{\rm tr}\Big[{\mathscr G}(k_1)
 \partial_\nu {\mathscr G}^{-1}\left(\bar{\bf k}\right)  {\cal D}_a{\mathscr G}(k_2)\partial_\mu {\mathscr G}^{-1}\left(\bar{\bf k}\right){\cal D}_b\Big],
\end{split}\\
& &
\begin{split}
S^{(2),(2)}_{\rm eff}=&{1 \over 2}{\rm tr}\left(\sigma^a \sigma^b\right)\sum_{{\bf k} {\bf k}'}\int _{\omega,\omega'}
{\cal A}^\nu_a(k)
{\cal A}^\mu_b(-k)\\
& \times{\rm tr}\Big[{\mathscr G}(k')\partial_\nu \partial_\mu {\mathscr G}^{-1}({\bf k}'){\cal D}_a {\cal D}_a\Big],
\end{split}
\end{eqnarray}
where $\bar{\bf k}\equiv \frac{{\bf k}_1+{\bf k}_2}{2}$, and we have once again made use of identity~\eqref{eq:trace}. For the simplicity of notation, it is implied that the derivatives only act on a single proceeding term, i.e., $\partial_\nu {\mathscr G}^{-1}\left(\bar{\bf k}\right)\dots$ is shorthand for $\left[\partial_\nu {\mathscr G}^{-1}\left(\bar{\bf k}\right)\right]\dots$. By a simple change of variables, and using ${\rm tr}\left(\sigma^a \sigma^b\right)=2\delta_{ab}$, we can write $S^{(2),(1)}_{\rm eff}$ as
 \begin{equation}\label{eq:s_eff_21}
\begin{split}
&S^{(2),(1)}_{\rm eff}=\sum_{{\bf k} {\bf k}'}\int_{\omega ,\omega'}{\cal A}^\nu_a(k){\cal A}^\mu_a(-k)\\
&\times{\rm tr}\Bigg[{\mathscr G}(k'+k)\partial_\nu {\mathscr G}^{-1}\left({\bf k}'+{{\bf k}\over 2}\right)  {\cal D}_a{\mathscr G}(k')\partial_\mu {\mathscr G}^{-1}\left({\bf k}'+{{\bf k}\over 2}\right){\cal D}_a\Bigg].
\end{split}
\end{equation}
%While ${\bf k}'={\bf k}_2$ runs over the entire Brillouin zone, the summation range for momentum ${\bf k}={\bf k}_1-{\bf k}_2$ depends on ${\bf k}'$ (is the Brillouin zone shifted by ${\bf k}'$). The problem is however superficial as explained below. Consider variables $x_1$, and $x_2$ running from $0$ to $2\pi$. Then $x'=x_2$ also runs from $0$ to $2\pi$, and $x=x_1-x_2$ runs from $-x'$ to $2\pi-x'$, which can be shifted to a Brillouin zone.
% 
As we are interested in the long-distance behavior, we can expand the trace in the expression above in $\bf k$. To zeroth order, we obtain a contribution to ${\cal C}^{0}_{ab,\nu \mu}$, while the first order term contributes to  ${\cal C}^{1}_{ab,\eta\nu \mu}$ [see Eq.~\eqref{eq:struct}]. The term $S^{(2),(2)}_{\rm eff}$, on the other hand has no dependence on $k$ in the trace, and therefore only contributes to ${\cal C}^{0}_{ab,\nu \mu}$.

\textit{Mass terms: }Let us first focus on the mass-term coefficient  ${\cal C}^{0}_{ab,\nu \mu}$. The discussion above implies that
 \begin{equation}\label{eq:s_eff_22}
{\cal C}^{0}_{ab,\nu \mu}=\delta_{ab} {\cal C}^{0}_{a,\nu \mu}, \qquad  {\cal C}^{0}_{a,\nu \mu}= {\cal C}^{(1),0}_{a,\nu \mu}+ {\cal C}^{(2),0}_{a,\nu \mu},
\end{equation}
where ${\cal C}^{(1),0}_{a,\nu \mu}$ and ${\cal C}^{(2),0}_{a,\nu \mu}$ are respectively the contributions of $S^{(2),(1)}_{\rm eff}$ and $S^{(2),(2)}_{\rm eff}$, explicitly given by
 \begin{eqnarray}\label{eq:C210}
{\cal C}_{a,\nu \mu}^{(1),0}&=&\sum_{{\bf k} }\int_{\omega}
{\rm tr}\Big[{\mathscr G}(k)
 \partial_\nu {\mathscr G}^{-1}\left({\bf k}\right)  {\cal D}_a{\mathscr G}(k)\partial_\mu {\mathscr G}^{-1}\left({\bf k}\right){\cal D}_a\Big],\\
 {\cal C}_{a,\nu \mu}^{(2),0}&=&\sum_{{\bf k} }\int_{\omega}{\rm tr}\left[{\mathscr G}(k)\partial_\nu \partial_\mu {\mathscr G}^{-1}({\bf k}){\cal D}_a {\cal D}_a\right],\label{eq:C220}
\end{eqnarray}
Using ${\cal D}_a^2=\openone$ and integration by parts, we can the write
 \begin{equation}
{\cal C}_{a,\nu \mu}^{(2),0}=-\sum_{{\bf k} }\int_{\omega}{\rm tr}\left[{\mathscr G}(k)\partial_\nu {\mathscr G}^{-1}({\bf k}){\mathscr G}(k)\partial_\mu {\mathscr G}^{-1}({\bf k})\right].
\end{equation}
Comparing with Eq.~\eqref{eq:C210} indicates that for $a=0$ (${\cal D}_0=\openone$), the two terms cancel out and the coefficient of $ {\cal A}_0^\mu {\cal A}_0^\nu$ in the effective action vanishes as expected (for instance nonzero ${\cal A}_0^\nu{\cal A}_0^\nu$ would correspond to superconducting response). The coefficient of the mass term is then given by
 \begin{equation}\label{eq:s_eff_22}
{\cal C}^{0}_{a,\nu \mu}= {\cal C}^{(1),0}_{a,\nu \mu}- {\cal C}^{(1),0}_{0,\nu \mu}.
\end{equation}
 We show these coefficients in Fig.~\ref{fig:c2_0} as a function of $J/t$ (for details of the calculation, see Appendix.~\ref{app_b2}). Other ${\cal C}^{0}_{a,\nu \mu}$ coefficients not shown in Fig.~\ref{fig:c2_0}  vanish.

The nonvanishing ${\cal C}^{0}_{a,\nu \mu}$ terms above describe the energetics of twisting the magnetic texture around the $a$ axis (as mentioned before due to the absence of spin-orbit coupling the coordinate system for spin components is independent of the coordinate system of the real-space lattice). Consider a region in real space where every local moment is rotated around the $a$ axis. If this rotation is uniform then all the spins are rotated together and there is no energetic cost with respect to the ground state. However, there is a stiffness against nonuniform rotations. Consider an example texture, where as we move in the $x$ direction, the rotation angle of the spins around the fixed axis $a$ linearly increases. This leads to a constant nonvanishing ${\cal A}_a^x$ proportional to the rate of change $\partial_x \phi$ of the rotation angle in the $x$ direction. The excess energy density (with respect to the ground state energy) required for such twist then goes as $\propto {\cal C}^{0}_{a,xx}\left(\partial_x\phi\right)^2$.

For a more general twist,  if we decompose the rotations into rotations around three orthogonal axes, there is no cross term and the energies add up (due to the $\delta_{ab}$ in ${\cal C}^{0}_{ab,\nu \mu}$). If the rotation angle changes with both $x$ and $y$, however, there are cross terms ${\cal C}^{(2),0}_{a,xy}\partial_x\phi\partial_y\phi$, which is expected as the underlying lattice is not symmetric with respect to the $x$ and $y$ directions. The above stiffness coefficient can be used to study the energetics of distorted configurations of interest (e.g., a state with several vortices) and compute the force between different vortices, or the force between a vortex and a boundary. The terms proportional to ${\cal C}^{0}_{a,\tau\tau}$ describe the work we need to perform to rotate the whole spin texture around the $a$ axis with a constant angular velocity (spin stiffness against time-dependent rotations).

\begin{figure}
 \includegraphics[width=8cm]{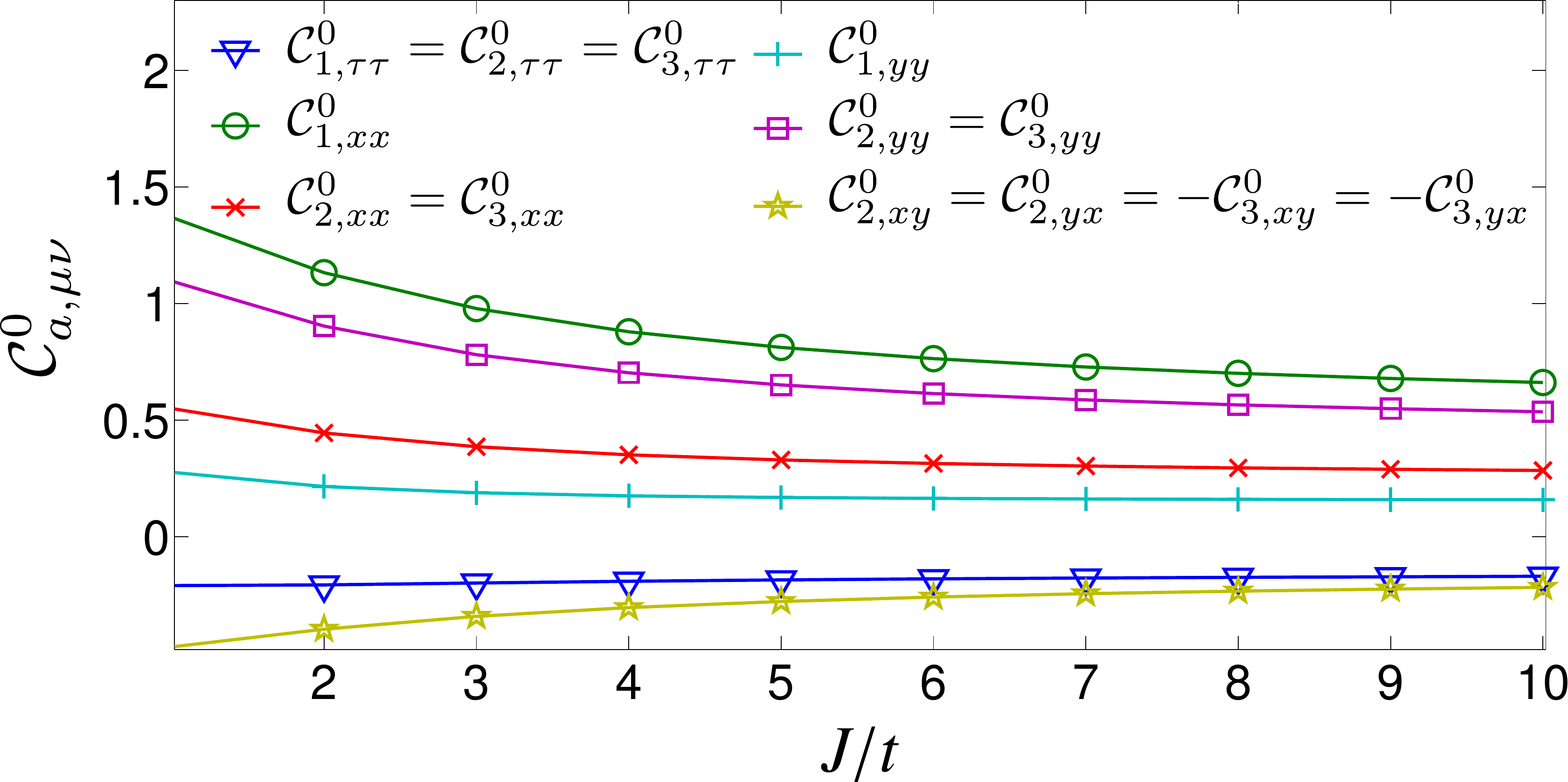}
\caption[]{\label{fig:c2_0}The nonvanishing coefficients of the mass term as a function of $J/t$. These coefficients saturate to constants for large $J/t$.}
\end{figure}
 
\begin{figure}
 \includegraphics[width=8cm]{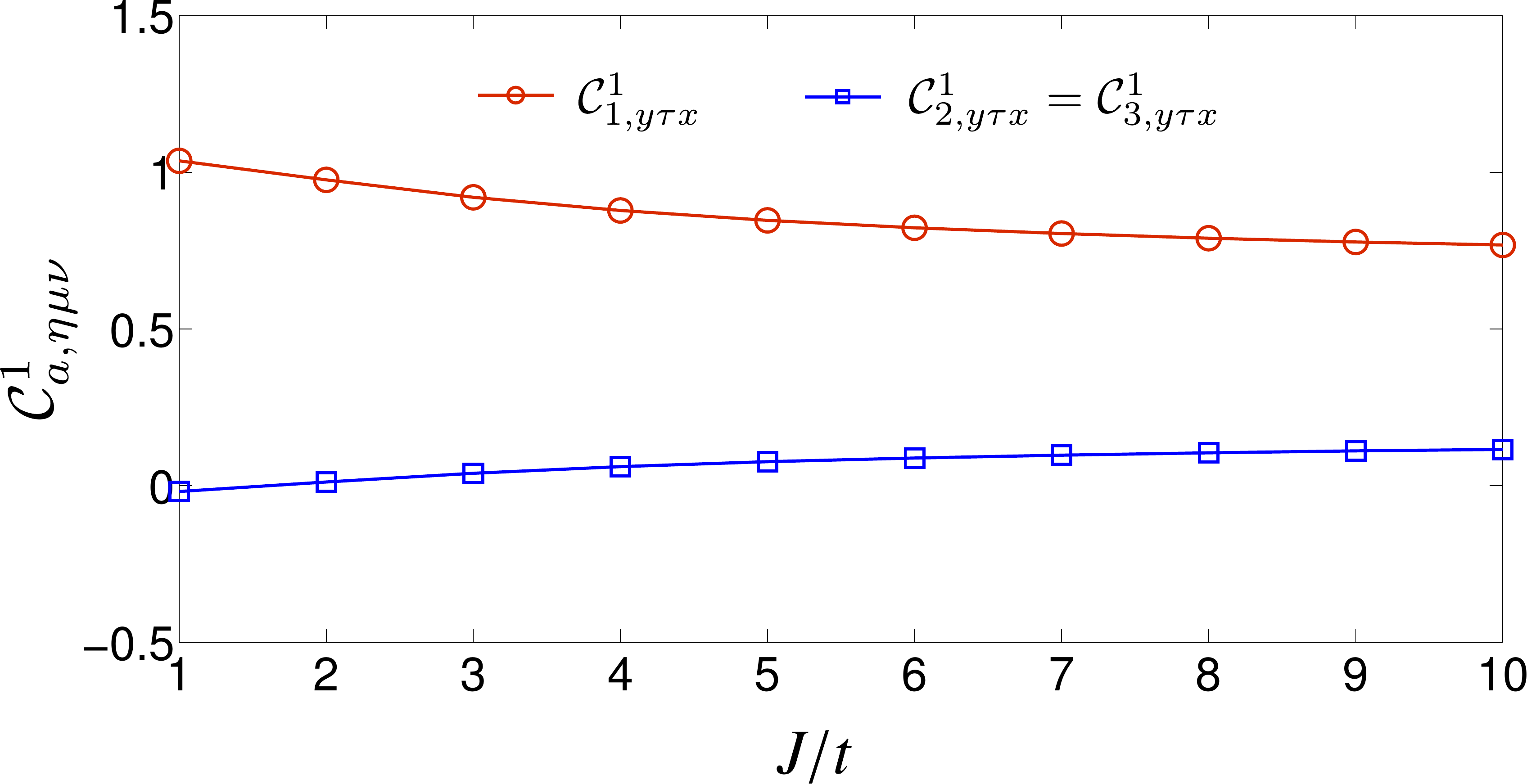}
\caption[]{\label{fig:c2_1}The nonvanishing coefficients ${\cal C}_{a,201}^{1}$ as a function of $J/t$, which determine all other nonquantized nonzero ${\cal C}_{a,\eta\nu \mu}^{1}$.}
\end{figure}

\textit{Response terms: } Thus far we have computed the nonquantized coefficients of the mass terms in Eq.~\eqref{eq:s_eff_22}, which encode the energetic cost (to second order in perturbation theory) of a twisted magnetic texture. We now turn to the
 \begin{equation}
{\cal C}_{ab,\eta\nu \mu}^{1} =\delta_{ab}{\cal C}_{a,\eta\nu \mu}^{1} 
\end{equation}
coefficients [see Eq.~\eqref{eq:struct}], which originate from expanding Eq.~\eqref{eq:s_eff_21} to the first order in $\bf k$. As we will see, these terms encode different quantized and nonquantized quantum-Hall response functions in charge and spin sectors. We start by expanding the trace in Eq.~\eqref{eq:s_eff_21} in $\bf k$ (and $\omega$). There are three terms that depend on $\bf k$: ${\mathscr G}(k'+k)$, $\partial_\nu {\mathscr G}^{-1}\left({\bf k}'+{{\bf k}\over 2}\right)$, and $\partial_\mu {\mathscr G}^{-1}\left({\bf k}'+{{\bf k}\over 2}\right)$. However, using the cyclic property of the trace and transformations $\mu\leftrightarrow\nu$ and ${\bf k} \leftrightarrow -{\bf k}$, we find that the contributions from expanding the last two terms cancel out. We then consider the expansion of ${\mathscr G}(k'+k)$, which gives a term 
\[
 {\mathscr G}(k+k')- {\mathscr G}(k')\approx - k_\eta{\mathscr G}(k') \partial_\eta {\mathscr G}^{-1}(k'){\mathscr G}(k').
\]
We can then write
\begin{equation}\label{eq:C211_1}
\begin{split}
{\cal C}_{a,\eta\nu \mu}^{1}=i\sum_{{\bf k} }&\int_{\omega}
{\rm tr}\Big[{\mathscr G}(k) \partial_\eta {\mathscr G}^{-1}(k){\mathscr G}(k)\\
&
\times \partial_\nu {\mathscr G}^{-1}\left(k\right)  {\cal D}_a{\mathscr G}(k)\partial_\mu {\mathscr G}^{-1}\left(k\right){\cal D}_a\Big].
\end{split}
\end{equation}
The above integral can be computed as explained in Appendix.~\ref{app_b3}. It turns out, as expected from the integer-quantum-Hall response of this system, that 
\begin{equation}
{\cal C}^{1}_{0, \eta\mu\nu}=-\epsilon_{\eta\mu\nu}.
\end{equation}
Moreover, we have the following quantized coefficients ${\cal C}^{1}_{a, 012}$ in the spin sector:
\begin{equation}
{\cal C}^{1}_{a, 012}=-{\cal C}^{1}_{a, 021}={1\over 3}, \qquad a=1,2,3,
\end{equation}
which encode the transverse spin $a$ current response to a twist of the magnetic texture around the $a$ axis.
Interestingly, in the spin sector $a=1,2, 3$, we obtain nonquantized (dependent on $J/t$) coefficients when the index $\eta$ in ${\cal C}^{1}_{a, \eta \nu \mu}$ is nonzero (corresponding to position derivative of the field ${\cal A}^\nu_a$). We have the following relationships between these nonquantized coefficients:
\begin{equation}
\begin{split}
&{\cal C}^{1}_{a, 120}=-{\cal C}^{1}_{a, 102},\qquad {\cal C}^{1}_{a, 201}=-{\cal C}^{1}_{a, 210}, \\
&{\cal C}^{1}_{a, 201}+{\cal C}^{1}_{a, 120}={2\over 3}, \qquad a=1,2,3.
\end{split}
\end{equation}
Given the relationships above, we just need to specify ${\cal C}^{1}_{a, 201}$ for $a=1,2,3$ as a function of $J/t$. The results are summarized in Fig.~\ref{fig:c2_1}. Notice that if a coefficient ${\cal C}^{1}_{a, \eta\nu\mu}$ (or its real or imaginary part) is symmetric in indices $\nu$ and $\mu$ (as opposed to antisymmetric), they cancel out in the effective action. In listing the above nonvanishing coefficients, we have taken this cancellation into account.

The above coefficients lead to several interesting electronic spin responses. From the antisymmetry in $\mu$ and $\nu$, we can explicitly write
\begin{equation}
\begin{split}
{\cal L}^1=&{\cal C}^{1}_{a,012}\left(\partial_0 {\cal A}^1_a {\cal A}^2_a-\partial_0 {\cal A}^2_a{\cal A}^1_a\right)+{\cal C}^{1}_{a,102}\left(\partial_1 {\cal A}^0_a {\cal A}^2_a-\partial_1{\cal A}^2_a{\cal A}^0_a\right)\\
&+{\cal C}^{1}_{a,201}\left(\partial_2 {\cal A}^0_a {\cal A}^1_a-\partial_2{\cal A}^1_a{\cal A}^0_a\right).
\end{split}
\end{equation}
Which leads to the following expectation values for spin currents:
\begin{eqnarray}
j_a^1&=&{\delta {\cal L}^1\over \delta {\cal A}^1_a}=-2{\cal C}^{1}_{a,012}\partial_0  {\cal A}^2_a+2{\cal C}^{1}_{a,201}\partial_2  {\cal A}^0_a,\\
j_a^2&=&{\delta {\cal L}^1\over \delta {\cal A}^1_a}=2{\cal C}^{1}_{a,012}\partial_0  {\cal A}^1_a+2{\cal C}^{1}_{a,102}\partial_1 {\cal A}^0_a.
\end{eqnarray}
The above expressions imply that making a time-dependent twist of the magnetic texture (around the $a$ axis) in the $x$ ($y$) direction leads to an expectation value for the spin-$a$ current in the $y$ ($x$) direction. Moreover if the velocity of spin rotation changes as we move in some real-space direction, we also get a contribution to the expectation value of the spin current. 

\section{Conclusions}\label{sec:conclusion}

In this paper, using an effective-action approach, we studied smooth distortions of noncoplanar magnetic textures with interactions mediated by itinerant electrons. We argued that parameterizing such distortions in terms of certain non-Abelian fields allows us to simultaneously study the energetics of the distortions as well as the responses of electrons to such twists. The electromagnetic filed can also be readily included in our formalism. Our work constitutes a first step toward a complete theory of low-energy magnetic excitations of noncoplanar textures coupled to itinerant electrons.

The action we obtain has some mass terms, which fully characterize the energetics of twisting the magnetic texture. Moreover, it yields the expectation values of charge and spin currents for insulator electronic states coupled to the magnetic texture. Topologically stable vortices are expected to play an important role in chiral magnets, and our action can serve as a basis for studying the energetics of a collection of vortices. In particular, by providing an energy functional, our action makes it possible to apply Langevin-type simulations to study the relaxation of magnetic textures. 

An extension of our formalism to include the required number of fast fluctuating modes (five modes in case of the triangular lattice with tetrahedral ordering), is expected to give rise to additive mass-like terms in the action characterizing the energetics of these modes (analogous to the ferromagnetic fluctuations in an antiferromagnet). In the Wess-Zumino terms, which contribute to the spin dynamics, however, the fast modes play a more important role.

\acknowledgments
We thank G.-W. Chern, O. Motrunich, D. Mozyrsky, C. Nayak, Y. Nishida, and S. Ryu for enlightening discussions.
This work was supported by the U. S. Department of Energy under the LANL/LDRD program (A. R, I. M, and, R. M).
Work performed at Argonne National Laboratory (by I. M.) was supported by the U. S. Department of Energy, Office of Science,
Office of Basic Energy Sciences, under Contract No. DE-AC02-06CH11357. R. M. also thanks CNPq (Brazil) for financial support.  

\appendix
\section{UNITARY TRANSFORMATION FOR BLOCK DIAGONALIZING THE HAMILTONIAN }\label{app1}
The explicit form of the unitary transformation~\eqref{eq:unit} is given by
\begin{equation}\label{eq:U}
{\cal U}={1\over 2}
\left(\begin{array}{cccc}
\sigma^0 &
\sigma^3 &
\sigma^0 &
\sigma^3
   \\
\sigma^1 &
i \sigma^2 &
-\sigma^1 &
-i \sigma^2
\\
i \sigma^2 &
\sigma^1 &
i \sigma^2 &
\sigma^1
\\
\sigma^3&
\sigma^0 &
-\sigma^3&
-\sigma^0
\end{array} 
\right).
\end{equation}
It can be shown that for the Hamiltonian of Eq.~\eqref{eq:H+momentum}, we have ${\cal U} {\cal H}({\bf k}) {\cal U}^\dagger=\sigma^0\otimes {\mathscr H}({\bf k})$, where
\begin{equation}\label{eq:H_4}
{\mathscr H}({\mathbf k})=
\left(\begin{array}{cccc}
 \varepsilon_0({\mathbf k}) &
-iJ &
J &
J
   \\
i J &
 \varepsilon_1({\mathbf k})&
-J &
J
\\
J &
-J &
 \varepsilon_2({\mathbf k}) &
iJ
\\
J&
J &
-iJ&
\varepsilon_3({\mathbf k})
\end{array} 
\right),
\end{equation}
with
\begin{eqnarray}
\varepsilon_0({\mathbf k})&\equiv &  \epsilon_1({\mathbf k})+ \epsilon_2({\mathbf k})+ \epsilon_3({\mathbf k}),\\
\varepsilon_1({\mathbf k})&\equiv & -\epsilon_1({\mathbf k})- \epsilon_2({\mathbf k})+ \epsilon_3({\mathbf k}),\\
\varepsilon_2({\mathbf k})&\equiv & \epsilon_1({\mathbf k})- \epsilon_2({\mathbf k})-\epsilon_3({\mathbf k}),\\
\varepsilon_3({\mathbf k})&\equiv &-\epsilon_1({\mathbf k})+ \epsilon_2({\mathbf k})-\epsilon_3({\mathbf k}).
\end{eqnarray}
The properties~\eqref{eq:prop1} and \eqref{eq:prop2} can also be obtained by direct matrix multiplication.
\section{CALCULATION OF THE COEFFICIENTS}

\subsection{The first-order term}\label{app_b1}
To compute Eq.~\eqref{eq:s_eff1} for $a=0$, we use the method of Sec.~\ref{sec:int} to write ${\rm tr}\left[{\mathscr G}\partial_\nu {\mathscr G}^{-1} \right]=\sum_{m,n}\langle n_{\bf k}|{\mathscr G}|m_{\bf k}\rangle \langle m_{\bf k}|\partial_\nu {\mathscr G}^{-1}|n_{\bf k}\rangle$, where $|m_{\bf k}\rangle$ and $|n_{\bf k}\rangle$ are the eigenvectors of the $4\times 4$ Hamiltonian ${\mathscr H}({\bf k})$ with respective eigenvalues $\varepsilon^m_{\bf k}$ and $\varepsilon^n_{\bf k}$. We then obtain ${\rm tr}\left[{\mathscr G}\partial_\nu {\mathscr G}^{-1} \right]=\sum_n{1 \over i\omega-\varepsilon^n_{\bf k}} \langle n_{\bf k}|\partial_\nu {\mathscr G}^{-1}|n_{\bf k}\rangle$, where $\langle n_{\bf k}|\partial_\nu {\mathscr G}^{-1}|n_{\bf k}\rangle$ is independent of $\omega$. We can now perform the integral over $\omega$, which gives
\[
\int d \omega{1 \over {i\omega-\varepsilon^n_{\bf k}}}=-\pi\: {\rm sgn}(\varepsilon^n_{\bf k}).
\]
Using ${\rm sgn}(\epsilon)=1-2n_0(\epsilon)$, where $n_0(\epsilon)$ is the zero-temperature Fermi distribution function, we can write
\begin{equation}
S^{(1)}_{\rm eff}= \pi {\cal A}^\nu_0(0,0)\sum_{n,{\bf k}}\left(2n_0(\varepsilon^n_{\bf k})-1\right)\langle n_{\bf k}|\partial_\nu {\mathscr G}^{-1}|n_{\bf k}\rangle
\end{equation}
The derivatives of $\partial_\nu{\mathscr G}^{-1}$ are
\begin{equation}
\partial_0 {\mathscr G}^{-1}=\openone,\qquad \partial_1 {\mathscr G}^{-1}= {\mathscr J}_x({\bf k}),\qquad \partial_2 {\mathscr G}^{-1}= {\mathscr J}_y({\bf k}),
\end{equation}
with the current operators defined as $
{\mathscr J}_x({\bf k})\equiv \frac{\partial {\mathscr H}({\bf k})}{\partial k_x}$ and  ${\mathscr J}_y({\bf k})\equiv \frac{\partial {\mathscr H}({\bf k})}{\partial k_y}$.
It is easy to observe that the matrices ${\mathscr J}_x$ and ${\mathscr J}_y$ are traceless [from Eq.~\eqref{eq:H_4}], so the summation over momenta is simply proportional to the ground-state expectation value of currents for $\nu>0$, which vanish in the bulk. Setting $\nu=0$ leads to Eq.~\eqref{eq:s_eff1}.

\subsection{Mass terms}\label{app_b2}

To compute the coefficient ${\cal C}_{a,\nu \mu}^{1,0}$ of Eq.~\eqref{eq:C210}, we once again make use of the eigenstates of $\mathscr H$ as follows:
 \begin{equation}\label{eq:c210}
\begin{split}
&{\cal C}_{a,\nu \mu}^{1,0}= \sum_{{\bf k},nm }\int_\omega{1 \over \left(i \omega-\varepsilon^n_{\bf k}\right)\left(i \omega-\varepsilon^m_{\bf k}\right)}\\
&\langle n_{\bf k}|\partial_\nu {\mathscr G}^{-1}\left({\bf k}\right)  {\cal D}_a|m_{\bf k}\rangle  \langle m_{\bf k}|  \partial_\mu {\mathscr G}^{-1}\left({\bf k}\right){\cal D}_a| n_{\bf k} \rangle
\end{split}
\end{equation}
The integral over $\omega$ gives
 \begin{equation}
\begin{split} 
\int d\omega &{1 \over \left(i \omega-\varepsilon^n_{\bf k}\right)\left(i \omega-\varepsilon^m_{\bf k}\right)}=\\
&{2 \pi \over \varepsilon^m_{\bf k}-\varepsilon^n_{\bf k}}\left[\Theta(\varepsilon^n_{\bf k} )\Theta(- \varepsilon^m_{\bf k})-\Theta(-\varepsilon^n_{\bf k} )\Theta( \varepsilon^m_{\bf k})\right],
\end{split}
\end{equation}
where $\Theta$ is the step function. The coefficients can then be explicitly computed by integration over momentum. As these terms characterize the excess energy of twisting the magnetic texture, it is not surprising that contour integration leads to a the typical structure of variations of energy in second-order perturbation theory.

\subsection{Response terms}\label{app_b3}
The computation of Eq.~\eqref{eq:C211_1} follows a similar method. In terms of the eigenstates of $\mathscr H$, we can write
\begin{equation}
\begin{split}\label{eq:C211_2}
&{\cal C}_{a,\eta\nu \mu}^{1,1}=-\sum_{{\bf k},nml }
\int_\omega {1\over \left(i \omega-\varepsilon^n_{\bf k}\right)\left(i \omega-\varepsilon^m_{\bf k}\right)\left(i \omega-\varepsilon^l_{\bf k}\right)}\\
&\times\langle n_{\bf k}|\partial_\eta {\mathscr G}^{-1}\left({\bf k}\right) |m_{\bf k}\rangle  \langle m_{\bf k}|  \partial_\nu {\mathscr G}^{-1}\left({\bf k}\right){\cal D}_a| l_{\bf k} \rangle \langle l_{\bf k}|  \partial_\mu {\mathscr G}^{-1}\left({\bf k}\right){\cal D}_a| n_{\bf k} \rangle.
\end{split}
\end{equation}
Once again we can do the integral over $\omega$ through contour integration. The integral vanishes if the three energies ($\varepsilon^n_{\bf k}$, $\varepsilon^v_{\bf k}$ and $\varepsilon^l_{\bf k}$) have the same sign (as all the poles lie to one side of the real axis). If one of the three energies, say $\varepsilon^n_{\bf k}$, has a different sign than the other two, then the integral gives $-{\rm sgn}(\varepsilon^n_{\bf k}){2\pi \over \left(\varepsilon^n_{\bf k}-\varepsilon^m_{\bf k}\right)\left(\varepsilon^n_{\bf k}-\varepsilon^l_{\bf k}\right)}$, and similarly for cases where $\varepsilon^m_{\bf k}$ or $\varepsilon^l_{\bf k}$ have a different sign. The result can be written in terms of sums of products of step functions.

\bibliography{itinerant_magnets}{}
\end{document}